\documentclass[11pt]{article}

\usepackage{latexsym, amsmath, amssymb}
\usepackage{mathrsfs}
\usepackage{bbm}
\usepackage{epsfig}

\newcommand{\qql}{\textquotedblleft}

\newcommand{\be}{\begin{eqnarray*}}
\newcommand{\ee}{\end{eqnarray*}}
\newcommand{\bea}{\begin{eqnarray}}
\newcommand{\eea}{\end{eqnarray}}

\newcommand{\Ba}{\mathbf{a}}
\newcommand{\Bb}{\mathbf{b}}
\newcommand{\Be}{\mathbf{e}}
\newcommand{\Bf}{\mathbf{f}}

\newcommand{\Br}{\mathbf{r}}
\newcommand{\Bw}{\mathbf{w}}

\newcommand{\BF}{\mathbf{F}}

\newcommand{\BH}{\mathbf{H}}
\newcommand{\BM}{\mathbf{M}}

\newcommand{\BX}{\mathbf{X}}

\newcommand{\hw}{\hat w}
\newcommand{\Bhw}{\mathbf{\hw}}
\newcommand{\Hw}{\mathbf{\hw}}
\newcommand{\Bal}{\mathbf{\alpha}}

\newcommand{\HO}{\mathbf{\hat\Omega}}
\newcommand{\BO}{\mathbf{\Omega}}
\newcommand{\Hf}{\hat \Bf}
\newcommand{\HX}{\hat \BX}
\newcommand{\iHO}{\HO^{-1}}

\title{Second Order Risk}
\author{Peter Shepard\\\it MSCI Barra\footnote{The author is vice president and senior researcher at MSCI Barra; 2100 Milvia Street; Berkeley, CA 94704; peter.shepard@mscibarra.com.}}
\date{August 17, 2009}

\begin{document} \setlength{\unitlength}{1mm}

\begin{titlepage}

\maketitle
\thispagestyle{empty}

\begin{abstract} 

Managing a portfolio to a risk model can tilt the portfolio toward weaknesses of the model.  As a result, the optimized portfolio acquires downside exposure to uncertainty in the model itself, what we call \qql second order risk."  We propose a risk measure that accounts for this bias.  Studies of real portfolios, in asset-by-asset and factor model contexts, demonstrate that second order risk contributes significantly to realized volatility, and that the proposed measure accurately forecasts the out-of-sample behavior of optimized portfolios.

\end{abstract}

\end{titlepage}

\section{Introduction}
Classical finance assumes the markets to be like a game of chance: Although future events are uncertain, the distribution of these events is known.  We cannot predict how the dice will land, but we can calculate the odds of any given outcome with certainty.  We can expect to roll snake-eyes on average one time in 36, and the rules of the game do not change without warning.

Unfortunately, real financial markets do not behave like a game of chance: Market volatility is itself volatile; hot industries come and go; new companies are listed and others merge or go bankrupt.  Under even the most generous assumptions, our estimates of financial risk are uncertain, based on limited historical observation, extrapolated forward.

For a passively invested portfolio, the effect of such uncertainty is as likely to be good or bad. The total risk may be overforecast or underforecast, but taken on average these errors tend to wash out.  On the contrary, an optimized portfolio is more likely to be hurt by uncertainty than helped by it.  Constructing portfolios to minimize risk can make them safer, but at the cost of introducing an asymmetric exposure to \qql second order risk."  

In this paper, we explore a framework to quantify and forecast second order risk.  Exploring only its mildest sources, we demonstrate that the act of optimizing a portfolio to a risk measure can render that measure systematically inaccurate.  However, rather than abandon risk measurement or ignore its uncertainties, the framework shows that we may begin to account for second order risk as we do more familiar sources of uncertainty.  

To quote a former US Secretary of Defense:

\medskip
{\it \qql There are known knowns. These are things we know that we know. There are known unknowns. That is to say, there are things that we know we don't know. But there are also unknown unknowns. There are things we don't know we don't know."

\medskip
	-Donald Rumsfeld, February 12, 2002} \cite{Rumsfeld}

\medskip
Our aim is to bring some of the latter into the category of \qql known unknowns."  Correcting for these uncertainties in general leads to a more conservative view of risk.  However, there will always be weaknesses in our models and much we cannot anticipate \cite{Taleb}. 
Perhaps a fourth category of \qql unknown knowns" is the most dangerous: things we think we know, but don't.

\subsection{A Toy Model}

To see the cause of second order risk, and how it can be forecast, consider the following toy example: Between two assets with the same expected return, an active manager aims to minimize risk by investing in the asset with the smaller standard deviation.  In this example, investors are constrained to hold a single asset.  After observing the returns of the assets, the manager finds Asset 1 to have a standard deviation of 8\%, while Asset 2 has a standard deviation of 11\%.  Placing a bet on Asset 1, the active manager believes the portfolio to have a risk of 8\%.

Although the manager doesn't know it, the returns of both assets are drawn from the same distribution, with standard deviation of 10\%.  The true risk is 10\%, regardless of which asset was chosen, but {\it the active manager's strategy is more likely to make investments whose risk happens to be underforecast}.  Meanwhile, passive investors are just as likely to hold either asset.  Looking at the same data, a passive investor holding Asset 1 would underforecast risk, while an investor holding Asset 2 would overforecast risk, but with no bias toward either outcome.

\begin{figure}[htp]
\centering
\includegraphics[scale=.8]{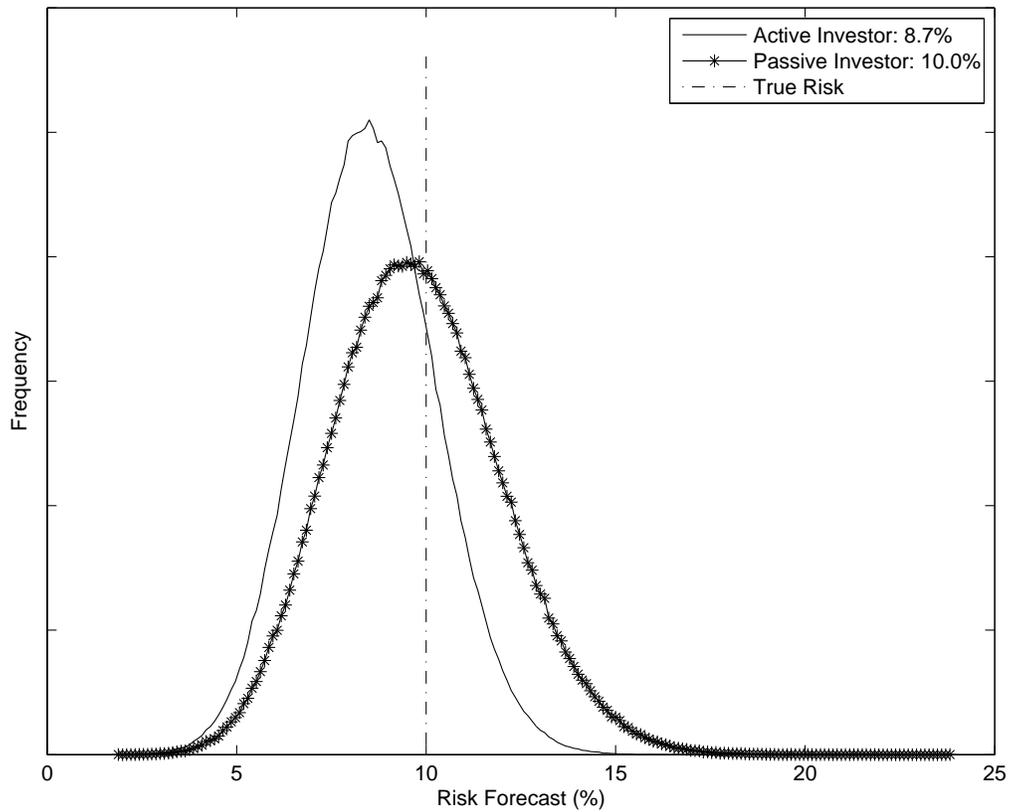}
\caption{The distribution of risk forecasts of a toy model of active and passive investors.  After observing the returns of two assets for ten periods, the active manager selects the asset with lower sample standard deviation, while a passive investor is equally likely to hold either asset.  Although the true risk is 10\% in all cases, the active manager consistently underforecasts risk.}
\label{fig:Toy}
\end{figure}

Figure \ref{fig:Toy} shows the result of repeating this experiment many times. In each trial, two time series are drawn from the same distribution and risk estimates are made.  The active manager bets on the asset with lower risk forecast, while the passive investor always holds Asset 1.  Noise diminishes the accuracy of both investors' risk forecasts, but it systematically biases only the active manager, whose average forecast is 8.7\%, less than the true 10\%.  The wise active manager would correct risk forecasts upward, to compensate for the bias introduced by active management.  Although the active manager does not know the true distribution of returns, we will see that it is possible to compensate for this bias.  


An intriguing implication is that the best risk forecast depends not just on the portfolio holdings, but also on the {\it strategy}.  In the simulation, the two managers hold identical portfolios in half of the trials, and forecast risk based on identical returns.  Nonetheless, because of differences of strategy, they have reason to make different risk forecasts, even when their portfolios exactly coincide.

\section{Model Uncertainty}

The example above is a case of aiming to maximize a utility function $U(\Bw)$ for which we have only an approximate model $\hat U(\Bw)$.  With perfect information, we would choose the variables $\Bw=\Bw^*$ to maximize $U(\Bw)$, but instead we must choose some other $\Hw$, the best guess given what is known.  The difference between $U(\Bw)$ and $\hat U(\Bw)$ leads to some discrepancy between the true best $\Bw^*$, and the best guess $\Hw$ given the available information.

\begin{figure}[htp]
\centering
\includegraphics[scale=.8]{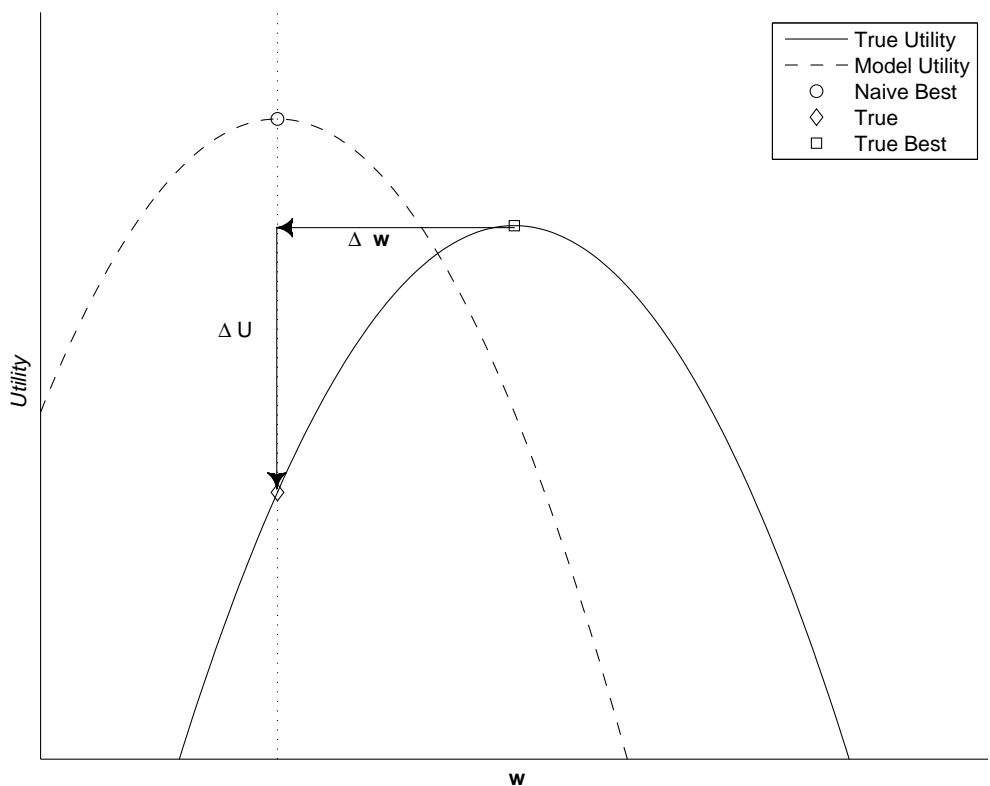}
\caption{The schematic effect of errors in the model utility function.  A model utility function usually has its maximum at least slightly removed from the true maximum, $\Delta \Hw$.  Though the realized $\Bw$ appears optimal to the model, it incurs a penalty $\Delta U$ under the true utility function.}
\label{fig:Util}
\end{figure}

As depicted in Figure \ref{fig:Util}, the effect of such a discrepancy is generically a loss: any departure $\Delta \Bw$ from  $\Bw^*$ reduces utility.  For small errors $\Delta \Bw$, the utility of $\Hw$ can be approximated
\bea
	U[\Hw] =  U[\Bw^*]+\Delta U \simeq U[\Bw^*]+\Delta \Bw' \mathbf{H} \Delta \Bw,
\eea 
where $\mathbf{H}$ is the Hessian of $U(\Bw)$ at $\Bw^*$, the matrix of second derivatives.  Simply because any function is concave at its maximum, $\BH$ is a negative-definite matrix, and the correction $\Delta \Bw' \mathbf{H} \Delta \Bw$ is negative for any $\Delta \Bw\neq 0$.  

Using the model $\hat U(\Bw)$ to forecast the utility of $\Hw$ misses the penalty $\Delta \Bw' \mathbf{H} \Delta \Bw$ that is the inevitable side-effect of having only an approximation.  If we can calculate a distribution of $\Delta \Bw$, we can account for the average loss $\sim E(\Delta \Bw' \mathbf{H} \Delta \Bw)$ that arises due to uncertainty in $\Bw^*$.

For the utility functions of finance, these errors are compounded by a tendency for models $\hat U(\Bw)$ to appear intrinsically better than the true $U(\Bw)$.\footnote{Physicists may recognize a relation to the tendency of quantum mechanical perturbations to systematically lower the ground state energy.}  As a result, the point labeled Naive Best in Figure \ref{fig:Util} is above even the True Best, attainable with perfect information utility $U$, and well above the True utility of $\Hw$.   
\medskip

\subsection{Uncertainty and Active Management}

Classical portfolio theory \cite{Markowitz} instructs the portfolio manager to build the optimal portfolio $\Bw$ from the covariance matrix $\BO$ and vector $\alpha$ of expected excess returns.  A variety of utility functions may be used, among them the Sharpe ratio
\bea
	U(\Bw)=\frac{\Bw'\alpha}{\sqrt{\Bw'\BO\Bw}}.  \label{SharpeRat}
\eea 
In the absence of constraints, the portfolio maximizing (\ref{SharpeRat}) has weights proportional to
\bea
	\Bw^*=\BO^{-1}\Bal. \label{wopt}
\eea
However, even assuming the markets to be stationary and Gaussian, the covariance matrix $\BO$ must be estimated from observation of historical behavior, which introduces noise.\footnote{Another source of noise, uncertainty in $\alpha$, may also be significant.  This subjective uncertainty could be incorporated into this framework, but we concentrate upon uncertainty in $\BO$, taking $\alpha$ to be known to the investor.}

Even if this noise level can be made relatively small, so that each element of $\BO$ is known with relative certainty, optimization tends to align the portfolio with the noise \cite{Michaud}, compounding many small errors into a large effect.  As the number of observations $T$ increases, the noise tends to be reduced by $\sim 1/T$, and a good estimator can insure that these errors average to zero.  

The impact of this small amount of noise is nonetheless significant.  For $\BO$ estimated directly from $N$ assets, we will see that the effect of noise on the optimized portfolio does not average to zero, but yields corrections of order 
\bea
\frac{1}{(1-N/T)}, \nonumber
\eea
growing without bound as the number of assets approaches the number of observations.  Since $T$ is limited by changing dynamics and market microstructure, this can lead to significantly inaccurate risk forecasts, and diminished out-of-sample performance.

For a factor model of risk \cite{Barr}, $N/T$ is replaced by the milder $K/T$, where $K$ is the number of systematic risk factors.  This makes portfolio optimization among many assets more robust to estimation errors, but may leave significant corrections to risk forecasts.

\subsection{A Second Order Risk Measure}
The denominator of the Sharpe ratio (\ref{SharpeRat}) is the standard deviation $\Sigma$ of future portfolio returns, a common measure of portfolio risk:
\bea
	 \Sigma^2=E_{\Br}\left((\Bw' \Br- \Bw'\bar \Br)^2 | \BO \right) &=& \Bw'\BO \Bw.	\label{PFvar}
\eea
Here  $E_x(f(x)|y)$ denotes an average over the variable $x$, conditional on $y$.\footnote{If there is no ambiguity, this may be denoted $E(f(x))$ or $E(f)$, to avoid cluttering the notation.}  If the true covariance matrix $\BO$ were known, (\ref{PFvar}) would be a good measure of uncertainty, but in practice we must make do with an estimate $\HO$ of the true distribution, based on observation.

Relative to the hypothetical true covariance matrix, the estimate $\HO$ is a random variable.  With $\HO$ used in place of $\BO$ in Equation (\ref{SharpeRat}), the optimized portfolio
\bea
	\Hw(\HO)=\HO^{-1}\alpha, \label{PF}
\eea
is also a random variable, distributed about the true optimal portfolio $\Bw^*$.  

The risk of $\Hw(\HO)$ therefore arises from two contributions.  In addition to the usual uncertainty of future returns $\Br$, there is a second risk associated with the randomness of the observation $\HO$ about $\BO$, which is typically neglected.  

To account for the latter uncertainty, we define a risk measure by extending the expectation value of Equation (\ref{PFvar}) to average over both ensembles:
\bea
	\Sigma^2_{SO} \equiv E_{\HO,\Br}\left( \left. (\Hw'\Br- \Hw'\Br)^2 \right| \BO \right).
\eea
Performing the average over the returns $\Br$ given $\HO$ we have
\bea
\Sigma^2_{SO} &=& E_\HO \left( \left. E_{\Br}\left( \left. (\Hw'\Br-\Hw'\bar\Br)^2 \right| \HO \right)\right| \BO \right)  \nonumber,
\eea
or
\bea 
\Sigma^2_{SO} &=& E_\HO \left( \left. \Hw' \BO \Hw \right| \BO \right). \label{therisk}
\eea
The final expression accounts for both the risk present in a given distribution and the additional risk due to distributional uncertainty.  Although similar in appearance to (\ref{PFvar}), it differs significantly in that it depends not on the portfolio holdings, but on the strategy that led to them, through $\Hw(\HO)$.  It is our aim to reliably estimate it.  

What is typically used to forecast risk, the \qql naive estimator" 
\bea
	 \hat \Sigma^2_{naive} = \Hw' \HO \Hw \label{naive},
\eea
may be significantly biased, even if the covariance matrix estimator $\HO$ is unbiased, $E_\HO ( \HO | \BO )=\BO$.  Active management induces a functional dependence $\Hw(\HO)$, a correlation between the portfolio and the estimation error in $\HO$, so that 
\bea
	E_\HO \left( \left. \Hw' \HO \Hw \right| \BO \right) & \neq & E_\HO \left( \left. \Hw' E_\HO(\HO|\BO) \Hw \right| \BO \right),
\eea
or
\bea 
E_\HO (  \hat \Sigma^2_{naive} | \BO ) \neq \Sigma^2_{SO}. \label{riskneq}
\eea 

The naive estimate of portfolio risk is typically lower than the true risk, and  lower even than the optimal risk attainable with perfect knowledge of $\BO$.  Intuitively, the optimized portfolio tends to overweight assets with underforecast risk, and to underweight assets whose risk $\HO$ overestimates.  

The degree of this bias grows with the uncertainty in $\HO$ and the sensitivity of the portfolio to $\HO$, via $\Hw(\HO)$.  For a portfolio constructed independent of $\HO$, such as a passive index fund, the left and right of (\ref{riskneq}) are equal.  

We compare $\Sigma^2_{SO}$ to the risk of the true, unknown optimal portfolio, $\Bw^*$.  Any portfolio on the efficient frontier is the minimum risk portfolio under a fixed return constraint.  For a minimum risk portfolio $\Hw$ subject to continuous constraints, we may formally expand the risk about $\Bw^*(\BO)$ as $\Hw=\Bw^*+\Delta \Hw$:
\bea
	E(\Hw' \BO \Hw)&=& E\left((\Bw^*+\Delta \Hw)' \BO (\Bw^*+\Delta \Hw)\right)  \nonumber\\
&=& \Bw^{*'}\BO \Bw^*+ 2\Bw^{*'}\BO E(\Delta \Hw)+E\left(\Delta \Hw'\BO\Delta \Hw\right)  \nonumber\\
&=& \Bw^{*'}\BO \Bw^{*}+E\left(\Delta \Hw'\BO\Delta \Hw\right).\label{riskdec}
\eea
The cross term $\Bw^{*'}\BO\Delta \Hw$ vanishes not just in expectation but for any $\Hw$ satisfying the constraints, by the optimality condition on $\Bw^*$.  The final expression (\ref{riskdec}) gives an intuitive decomposition of risk as that attainable with perfect knowledge of $\BO$ plus the cost of uncertainty.  

Note that $E\left(\Delta \Hw'\BO\Delta \Hw\right)$ is positive, so the effect of the uncertainty is a risk penalty.  Considering the portfolio $\Hw$ to be an estimator of the true optimal portfolio $\Bw^*$, Equation (\ref{riskdec}) quantifies the risk cost of estimation error.\footnote{bounded below by the Cramer-Rao bound of statistics.}

Although we focus on the uncertainty due to estimation errors in $\HO$, the expected value in Equation (\ref{therisk}) may also be extended to other sources of uncertainty, such as stochastic time-dependence in $\BO$ and $\alpha$.  To quantify this behavior requires additional modeling assumptions, resulting in greater subjectivity, but the result is qualitatively the same: optimization produces an asymmetric downside exposure to model uncertainty.

\section{Asset Covariance Matrix}\label{sec:AssetLevel}

We first explore second order risk in the context of the covariance matrix estimated directly from asset returns:
\bea
	\HO \equiv \frac{1}{T} \Br \Br'.
\eea
Here $\Br$ is the $N \times T$ matrix of de-meaned\footnote{For our purposes, neglecting the $\sim 1/T$ estimation error of ex-post mean returns is a harmless simplifying assumption, unrelated to the difficult question of quantifying uncertainty in the forecast $\alpha$.} returns of $N$ assets over $T$ observation periods.  Assuming Gaussian returns, $\HO$ follows a Wishart distribution \cite{Wishart}. 
 
For the simple portfolio of Equation (\ref{PF}), the risk of (\ref{therisk}) can be calculated explicitly.  In terms of the observable $\Bhw' \HO \Bhw$, we find
\bea
	\Sigma^2_{SO}=E(\Bhw' \BO \Bhw)&\simeq& E(\Bhw' \HO \Bhw)\left(1-\frac{N}{T}\right)^{-2}.\label{real3}
\eea
Details of the calculation are given in the Appendix.  

The significance of Equation (\ref{real3}) is twofold: it demonstrates the scale of the bias, and immediately suggests how to correct it.  Equation (\ref{real3}) implies
\bea  
	\hat\Sigma^2_{SO} \equiv \Bhw' \HO \Bhw \left(1-\frac{N}{T}\right)^{-2}  \label{AssetEstimator}
\eea
is an unbiased estimator of the risk of the optimized portfolio:
\bea
	E_\HO(\hat\Sigma^2_{SO}|\BO)=\Sigma^2_{SO}. \nonumber
\eea
Crucially, the correction for second order risk is a function of $N$ and $T$ only, which are known to the investor without additional information about $\BO$, making it possible to forecast second order risk.  For an investment universe of 500 assets, and an asset covariance matrix estimated from 4 years of daily returns, Equation (\ref{AssetEstimator}) doubles the predicted standard deviation of portfolio returns.


\subsection{Empirical Results}

\begin{figure}[htp]
\centering
\includegraphics[scale=.8]{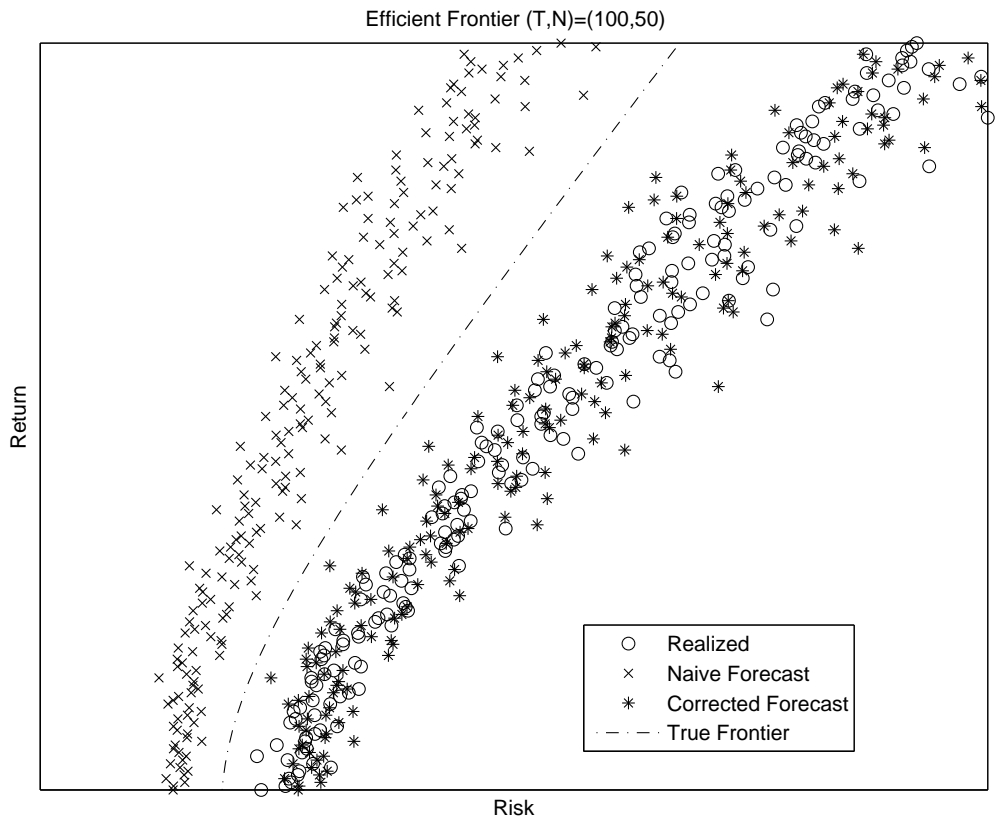}
\caption{Monte Carlo simulation. The average Corrected Forecast is the same as the average Realized risk, below the true efficient frontier accessible only with perfect information about $\BO$.  The Naive Forecast appears better than even the true efficient frontier. }
\label{fig:Frontier}
\end{figure}

The simplicity of the Sharpe ratio optimized portfolio (\ref{PF}) aided in deriving the simple second order risk correction in Equation (\ref{AssetEstimator}), but the inflation factor $\left(1-\frac{N}{T}\right)^{-2}$ can be a sufficient approximation to the correction needed for other utility functions.  

Figure \ref{fig:Frontier} shows the results of a Monte Carlo simulation for the minimum risk portfolio, constrained to be fully invested and to have fixed expected return $\Hw'\alpha=R$ with respect to a randomly chosen $\alpha$ and fixed covariance matrix $\BO$.  For each value of $R$, a new $\HO$ is estimated from $T=100$ observations of $N=50$ returns. 

The curve labeled \qql True Frontier" is the efficient frontier that could be achieved if $\BO$ were known with certainty, corresponding to the True Best point in Figure \ref{fig:Util}.  Risk along the true frontier is given by $\sqrt{\Bw^{*'}\BO \Bw^*}$.  

The points labeled \qql Realized" show the actual risk $\sqrt{\Hw '\BO \Hw}$ of the optimized portfolios, which correspond to True in Figure \ref{fig:Util}.  This risk is well above the optimal risk, showing that estimation error degrades performance by preventing the optimal hedging of risk.

The \qql Naive Forecast" risk, $\sqrt{\Hw '\HO \Hw}$, is seen to be significantly over-optimistic, on average by a factor of two.  Its location to the right of the true frontier is in correspondence with the position of the Naive Best point in Figure \ref{fig:Util}, overestimating not only the utility attainable with a model, but also what would be attainable with perfect information.

In contrast, the \qql Corrected Forecast" $\sqrt{\Hw '\HO \Hw}\left(1-\frac{N}{T}\right)^{-1}$ accurately captures the risk of the optimized portfolio.  Although its efficiency is diminished by noise, the corrected forecast provides unbiased estimates.

\begin{figure}[htp]
\centering
\includegraphics[scale=.9]{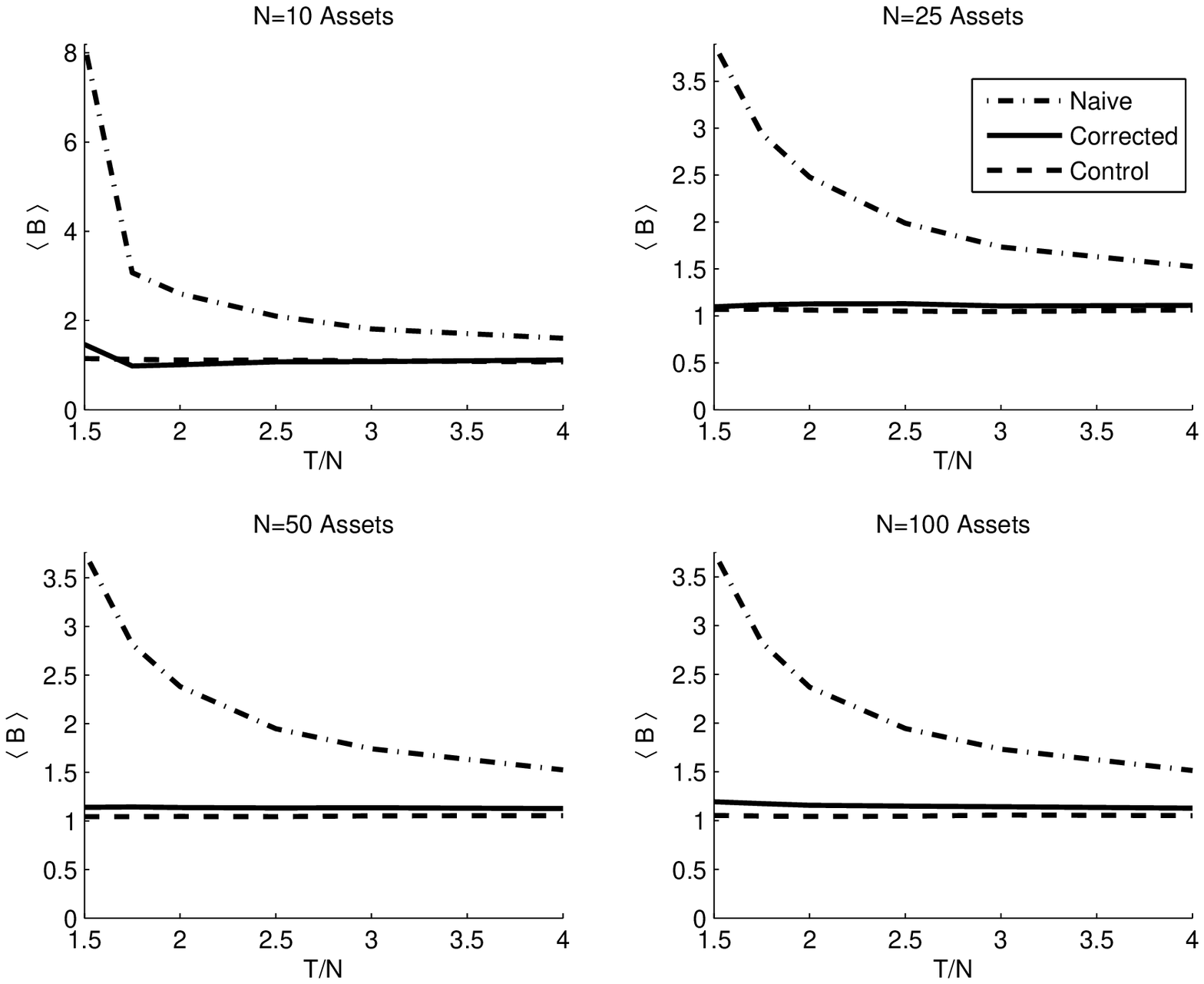}
\caption{Average bias statistics of real portfolios with varying number of assets $N$ and length of observation window $T$.  A bias statistic greater (less) than 1 indicates under- (over-) forecast risk.  \qql Naive" denotes the conventional risk forecast for the optimized portfolio, Equation (\ref{PF}), with random alpha vector among $N$ assets.  The \qql Corrected" forecasts are for the same portfolios as Naive, but with forecasts corrected for second order risk.  The bias statistics of random portfolios among the same assets are shown for comparison.}
\label{fig:BiasStats2}
\end{figure}

Testing the methodology with real market data is complicated by the fact that the \qql true" covariance matrix is not known, so $\Hw' \BO \Hw$ must be estimated by observing realized volatility.  In the context of portfolios, forecasts, and market conditions that are changing in time, the Bias Statistic is a useful tool for testing the accuracy of risk forecasts.  

For a given time series of portfolios $\Hw_t$, the bias statistic is constructed from the forecast standard deviations $\hat\Sigma_t$ and out-of-sample realized returns $R_{t+1}$ as
\bea
	B \equiv std \left(\frac{R_{t+1}}{\hat\Sigma_t}\right).    \label{BiasStatistic}
\eea
An underforecast $\hat\Sigma_t$ produces bias statistics greater than 1, while overforecasts lead to $B<1$.  

We construct portfolios from a universe of the largest stocks in the United States, studying daily returns for the ten years through April 2009.  To reduce the effects of extreme events, asset returns are trimmed to -50\% and +80\%.  Assets with an incomplete history are excluded from the sample, leaving about 1800 stocks.

For each trial, we construct portfolios among $N$ stocks selected at random.  Optimized portfolios are constructed according to Equation (\ref{PF}), with $\HO$ estimated from a rolling window of $T$ days, and a random $\alpha$ vector.  Each day, risk forecasts are constructed with the naive estimator (\ref{naive}) and corrected estimator (\ref{AssetEstimator}), from which bias statistics are calculated.  As a control, we also construct random portfolios from the same stocks, which are not subject to the bias of second order risk.   

Figure \ref{fig:BiasStats2} shows the average bias statistics over 50 trials, for portfolios of $N$=10, 25, 50 and 100 assets, and observation windows $T$ with $T/N$=1.5, 1.75, 2, 2.5, 3 and 4.  In all cases, the standard risk forecast of optimized portfolios significantly underforecasts realized volatility, as indicated by bias statistics greater than 1.  Comparison with the control demonstrates that the underforecast risk is a result of optimization, not some other feature of the distribution.  

Second order risk is therefore responsible for a significant portion of the out of sample portfolio volatility.  In contrast, the corrected forecasts are nearly as accurate as the control, confirming the validity of the estimator $\hat \Sigma^2_{SO}$.

\section{Factor Process} \label{sec:Factor}

Section \ref{sec:AssetLevel} assumed the returns generating process to be a multivariate Gaussian at the level of the assets, with no additional structure.  As the resulting estimation error effects went like $N/T$, a large investment universe may require decades of data for robust optimization, far longer than the timescales over which market relationships are stable.

Perhaps due to experience with such effects, most practitioners do not build large optimized portfolios from covariance matrices estimated directly from the assets, instead using more robust factor models.  Rather than estimate the correlation among every pair of assets, a smaller number of systematic factors is identified, such as Value, Momentum, or industry membership.  The return of each asset is decomposed into the $K \times T$ factor return $\Hf$ and the asset-specific, or idiosyncratic, return~$\hat\Be$.  
\bea
	\Br=\HX\Hf+\hat\Be.   \nonumber
\eea
The $N\times K$ matrix $\HX$ defines the model exposure of each asset to the $K$ factors.  It is assumed that all correlation among assets is due to their common exposure to the factors, so that the covariance matrix takes the form
\bea
	\HO=\HX \hat \BF \HX'+\hat \Delta,   \label{OmegaF}
\eea
where the estimated factor covariance matrix is defined to be $\hat \BF \equiv \Hf \cdot \Hf'/T$,\footnote{We again neglect ex-post mean returns, for simplicity of exposition.} and the specific risk matrix $\hat\Delta$ is assumed to be diagonal.  For a broadly diversified portfolio of $N$ assets, the specific risk contribution is suppressed by a factor of $1/N$ relative to the factor risk, assuming the latter has not been hedged away.  Because it is diagonal, estimation error in the specific risk matrix does not have the effects of covariance matrices, as an optimizer is not fooled into hedging out spurious correlations.  For simplicity, we use the true $\Delta$ rather than an estimate, which omits corrections suppressed to order $1/N$ compared to the effects under study.

Rather than estimate all $N(N+1)/2$ elements of the asset covariance matrix, the smaller $K\times K$ factor covariance matrix can be estimated from a much shorter time history, allowing a better reflection of the current market conditions.  Provided the factor model accurately captures the returns process, it yields a far more robust estimate of the covariance among assets.

\subsection{Factor Modeling Errors}\label{sec:ModelError}
Unfortunately, factor models may be subject to a variety of errors, which can degrade their performance.  Similar to the asset-level case above, the factor covariance matrix has estimation errors due to the finite number of observations from which it is estimated.  These errors are of order $K/T$,  typically far better than the $N/T$ behavior of errors of the asset covariance matrix, but significant.

Unlike the asset-level case, factor models have an additional source of error in the factor exposures $\HX$, which is more difficult to quantify.  In general, determining the exposures is an inexact science, requiring financial insight as much as straightforward econometric technique.  Errors in the exposures, though difficult to measure, can skew the risk forecasts of portfolios, particularly those constructed to minimize exposure to systematic risk.

Two classes of exposures errors may be identified, which we call \qql coherent" and \qql incoherent."  Incoherent errors are those that are uncorrelated with the portfolio, so that their aggregate effect is diversified.  For example, if the true exposures $\BX$ differ from the model exposures by uncorrelated random noise, $\BX=\HX+\epsilon$, then the portfolio exposure error is approximately
\bea
	\sum_i w_i \epsilon_i &\sim & \frac{1}{N} \sqrt{N} std(\epsilon) \nonumber \\
	&\sim & \frac{std(\epsilon)}{\sqrt{N}}, \nonumber
\eea
giving a contribution to variance $\sim 1/N$, behaving like diversifiable specific risk rather than systematic factor risk.  The $\sqrt{N}$ suppression is due to the low likelihood that many terms contribute with the same sign.

A more dangerous class of errors, the coherent exposure errors, are those for which the portfolio is likely to accumulate a finite exposure.  These errors can be generated by discrepancies between the alpha signal and the risk model, such as small differences in factor definitions.  Similar to other errors we have seen, the optimized portfolio tends to align itself with these errors, resulting in an unsuppressed, non-diversifiable contribution
\bea
	\sum_i w_i \epsilon_i &\sim & \frac{1}{N} N std(\epsilon) \nonumber\\
	&\sim & std(\epsilon). \nonumber
\eea
For example, if a risk model defines a Value factor in terms of the Book-to-Price ratio, while the alpha Value factor uses Earnings-to-Price, the optimized portfolio may make large unintended bets \cite{Dan} on the difference between these Value factor definitions.  The contribution from this hidden exposure may be significant for a portfolio that has hedged away known factor exposures.

To see how these coherent exposure errors occur, consider the estimated maximum Sharpe ratio portfolio, Equation (\ref{naive}).  Without loss of generality, we work in a basis of assets such that the specific risk is uniform, $\Delta=\sigma^2 \mathbf{1}_N$.\footnote{This can be achieved by taking
$
\alpha_i \rightarrow \alpha_i \sigma/\sigma_i, 
r_i \rightarrow,  r_i \sigma/\sigma_i  \label{scaling}, 
X_i \rightarrow X_i \sigma/\sigma_i,
$
which is the usual map from optimal regression weights proportional to $\sigma_i^{-2}$ to an equivalent OLS regression.}  Furthermore, we choose a basis of model factors such that $\HX'\HX=N\mathbf{1}_K$.  The factor of $N$ captures the scaling behavior of $\HX'\HX$ with the number of assets.

For the factor model of Equation (\ref{OmegaF}), it is useful to decompose $\alpha$ into components in the plane of the model exposures, and an orthogonal piece
\bea
	\alpha &=&  \HX\Ba + \alpha_\bot, \label{alphadec}
\eea
where $\alpha_\bot  \equiv ({\mathbf 1}-\frac{\HX\HX'}{N}) \alpha$ satisfies $\HX'\alpha_\bot=0$, and $\Ba \equiv \frac{\HX'\alpha}{N}$.  After some algebra, Equations (\ref{PF}), (\ref{OmegaF}) and (\ref{alphadec}) yield
\bea
	\Hw = w \left(\alpha_\bot+ \sigma^{2} N^{-1}\HX \hat \BF^{-1}\Ba \right)+\dots,  \label{walpha}
\eea
where $w$ is an arbitrary constant.  The $\dots$ represent terms $\mathcal{O}\left((\frac{\sigma^2}{N} \hat\BF^{-1})^{2}\right)$, which are suppressed by a further factor of $N$ in a large universe of assets.  

Equation (\ref{walpha}) shows the potential for distortions from a misalignment of the alpha factors and risk factors, with $\mathcal{O}(N^{-1})$ suppression for the factor component of $\alpha$ relative to the component $\alpha_\bot$ orthogonal to the plane of known factor risk.  The model factor risk
\bea
	\Hw' \HX \hat \BF \HX' \Hw = \Ba' \hat\BF^{-1}\Ba \label{ModelF}
\eea
sees no systematic risk in $\alpha_\bot$.  

Unless the manager has extraordinary skill, $\alpha_\bot$ may represent noise rather than an arbitrage opportunity\footnote{Even managers using a stock-picking strategy, rather than making explicit factor bets, may be judging assets on a handful of criteria that constitute risk factors.} (in the sense of the Arbitrage Pricing Theory), and optimization points the portfolio into a blind spot of the risk model.

These coherent exposure errors are avoided if the factor exposures contain the full alpha signal,
\bea
	\alpha= \HX \Ba.  \label{AlphaX}
\eea
If the alpha signal contains a component orthogonal to the plane spanned by the factor exposures, this component can be used to estimate an additional factor \cite{Jose}:
\bea
	\Hf^{(\alpha_\bot)}= \frac{\alpha_\bot'\Br}{\alpha_\bot^2}.  \label{AlphaFactor}
\eea
Although the resulting exposures may still contain errors, they are of the safe, incoherent variety, as the portfolio is unlikely to align along them.

If it is not feasible to estimate an additional factor from the alpha signal, so that $\alpha$ is in the plane of model factor exposures, we do not know whether the orthogonal component $\alpha_\bot$ represents an arbitrage opportunity, or just a limitation of the risk model.  

To see this, assume that there are true risk factors $\BX$, which differ from the model factors $\HX$ by 
\bea
	\BX = \HX + \epsilon.  \label{modelnoise}
\eea
We can assume without loss of generality that the noise is orthogonal to the model exposures\footnote{Since factor processes are invariant under an arbitrary change of basis in the space of factors, $\BX \rightarrow \BX \BM$, if $\HX'\epsilon\neq 0$, we can redefine $\BX \rightarrow \BX \left(\HX'\BX\right)^{-1} \HX'\HX$, so that $\HX'\epsilon=\HX'(\BX-\HX)=0$.},
\bea
	\HX'\epsilon=0. \label{orthog}
\eea 
A bit of algebra shows that the estimated factor returns $\hat \Bf$ associated with $\HX$ are related to the true factor returns $\Bf$ as 
\bea
	\hat \Bf=\Bf+\HX'\Be/N.   \label{Fhat}
\eea
Therefore, at large $N$ and $T$, the estimated factor covariance matrix is accurate, $\hat \BF \simeq \BF$, and the true factor risk of (\ref{walpha}) is related to the model, Equation (\ref{ModelF}), by
\bea
	\Hw'\BX\BF\BX '\Hw = \Hw'\HX\hat \BF\HX '\Hw + w^2\left(\alpha_\bot' \epsilon \BF \epsilon' \alpha_\bot +2\sigma^2 \Ba'\epsilon'\alpha_\bot\right).   \label{trueF}
\eea 
The terms on the right represent a contribution to the factor risk that the model does not see, a form of second order risk.  The noise $\epsilon$ is unknown, but the orientation of the portfolio to the plane of model exposures provides insight into its magnitude.  

Assuming \qql no arbitrage", that $\alpha$ lies fully in the plane of the true factors $\BX$
\bea
	\alpha=\BX \Ba, \label{modelalpha}
\eea
then $\alpha_\bot=\epsilon \Ba$, 
and the remaining $\epsilon$ dependence is of the form $\epsilon' \epsilon$, which in a large universe is sensitive only to the statistics of $\epsilon$, rather than its details.  Imposing the further assumption that $\epsilon$ arises from white noise, 
\bea
\epsilon' \epsilon \simeq N\rho^2 \mathbf{1}_K, \label{epsilon}
\eea
where $\rho$ is an unknown noise parameter, we may solve for the total factor risk of (\ref{trueF})
\bea 
	\Hw'\BX\BF\BX '\Hw = \Hw'\HX\hat \BF\HX '\Hw + w^2 \left( \frac{\alpha_\bot^4 \Ba' \BF \Ba}{\Ba^4} + 2 \alpha_\bot^2 \sigma^2 \right). \label{Exp}
\eea 
Like Equation (\ref{AssetEstimator}), the right-hand side forecasts second order risk using only information available to the investor.  Though relying on the assumptions of Equations (\ref{modelalpha}) and (\ref{epsilon}), these corrections warn of the susceptibility to second order risk for a portfolio tilted far out of the plane of the model exposures.

\subsection{Factor Model Estimation Error}

If the model exposures contain $\alpha$, as in Equation (\ref{AlphaX}), then the  effects of modeling errors discussed in Section \ref{sec:ModelError} are replaced by the milder effects of estimation error.

We continue to assume as in Equations (\ref{modelnoise}) and (\ref{orthog}) that there is unknown noise in the exposures.  Equation (\ref{Fhat}) implies $\hat \Bf=\Bf$, at large $N$.  Unlike in Section \ref{sec:ModelError}, we do not assume the large $T$ limit, so there is estimation error in the model factor covariance matrix $\hat \BF=\frac{\Bf \cdot \Bf'}{T}$.  With the assumption $\alpha_\bot=0$, the true factor risk of the portfolio (\ref{walpha}) is 
\bea
	\Hw'\BX \BF \BX' \Hw &=& N^{-2} \Ba' \hat\BF^{-1}  \HX' \BX \BF \BX ' \HX \hat\BF^{-1} \Ba  \nonumber\\
	&=&  \Ba' \hat\BF^{-1}  \BF  \hat\BF^{-1} \Ba,  \label{Ftrue}
\eea
while the naive estimate is
\bea
	\Hw'\HX \hat \BF \HX' \Hw &=& \Ba' \hat\BF^{-1} \Ba.  \label{Fnaive}
\eea
In close analogy with the asset-level case, taking expected values of Equations (\ref{Ftrue}) and (\ref{Fnaive}) yields
\bea
	\Sigma^2_{SO,f} \equiv E_{\hat \BF}\left( \left.\Hw'\BX \BF \BX' \Hw \right| \BF \right)=\left(1 - \frac{K}{T}\right)^{-2} E_{\hat \BF}\left( \left.\Hw'\HX \hat\BF \HX' \Hw \right| \BF \right). \label{FactorBias}
\eea
Therefore, the estimator 
\bea
	\hat \Sigma^2_{SO,f} \equiv \left(1 - \frac{K}{T}\right)^{-2} \Hw'\HX \hat\BF \HX' \Hw \label{FactorRisk}
\eea
provides unbiased forecasts, without additional knowledge of the true factors. More complicated corrections hold beyond the $N \rightarrow \infty $ limit.  
Despite the true factor risk depending on both the unknown exposures $\BX$ and factor covariance matrix $\BF$, Equation (\ref{FactorRisk}) forecasts second order risk using only observable quantities.

While the $\frac{N}{T}$ effects of Section \ref{sec:AssetLevel} could overwhelm a large portfolio, these $\frac{K}{T}$ are more likely to be under control.  However, for a typical factor model of $K\sim 50$ factors and effective observation window\footnote{See the discussion of effective observation window in Section \ref{sec:Extensions}.} $T \sim 200$, the $\sim30\%$ boost to forecast volatility is an important correction.

\subsection{Empirical Results}

To study the effects of factor model estimation errors, we consider the Barra Global Equity Model (GEM2) \cite{GEM2}.  The model estimates a World factor, 34 industry factors, 55 country factors and 8 style factors from an estimation universe based on the MSCI All Country World Index, consisting of about 8000 stocks.  

We consider an ensemble of 500 portfolios constructed from the estimation universe, optimized using (\ref{walpha}) relative to $\alpha=\HX \Ba$, with random vector $\Ba$.\footnote{The two-fold exact multicollinearity of GEM2 is resolved by projecting to a smaller subspace of factors.}  To make contact with Equation (\ref{FactorRisk}) we consider $\HO$ given by Equation (\ref{OmegaF}), with $\hat \BF$ and $\hat \Delta$ estimated with equally weighted covariance estimators over a rolling window of $T=156$ weekly returns. We discuss the corrections necessary for the exponentially weighted, Newey-West estimator used in GEM2 in Section \ref{sec:Extensions}.  To reduce the effect of extreme events, asset returns outside of $(-80\%,400\%)$ or exceeding ten cross-sectional standard deviations are dropped.

As a control, we also construct 500 portfolios of the form $\Hw=\HX \Bb$, with $\Bb$ a random vector.  Though exposed to factor risk, these portfolios are not subject to the biased second order risk of optimized portfolios.

\begin{figure}[htp]
\centering
\includegraphics[scale=.7]{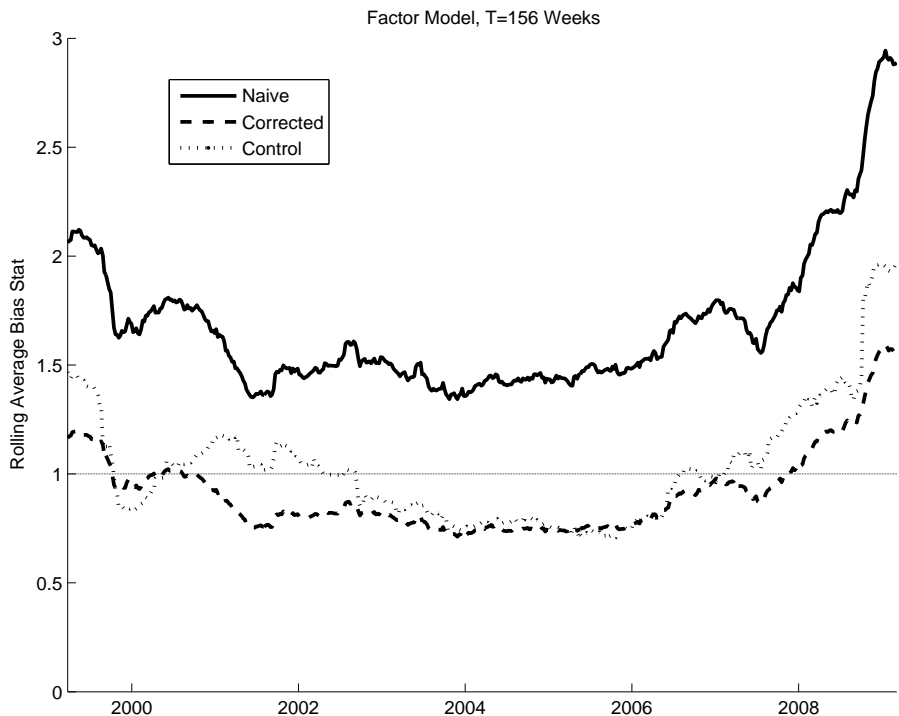}
\caption{Average trailing bias statistics for ensembles of 500 optimized and random portfolios.}
\label{fig:FactorBias}
\end{figure}

Each week, the risk is forecast using the standard risk estimator and the corrected estimator (\ref{FactorBias}).  Bias Statistics (\ref{BiasStatistic}) are calculated with a standard deviation over a trailing 52 week window, over the decade ending March 2009.

The results in Figure \ref{fig:FactorBias} are easily interpreted.  The control has bias statistics greater than one during periods of upward-trending volatility, prior to 2000 and since late 2007.  Similarly, the control has bias statistics consistently below one between 2001 and early 2007, during which volatility trended downward and the trailing window overforecast volatility of future returns.  The deviation from one is exacerbated by the use of the OLS estimator for this study, which is less responsive to changing market conditions.

The naive risk forecasts of the optimized portfolios have bias statistics significantly greater than one for the whole of the study, demonstrating that volatility is underforecast when second order risk is neglected.  Since the risk forecasts are so much worse than those of the control, the error must be due to optimization, rather than the underlying factor model.  An average bias statistic of about two indicates that the naive forecasts only capture about half the true risk.  The risk due to distributional uncertainty therefore contributes fully half of the risk of these portfolios.

In comparison, the corrected risk forecasts match the accuracy of the control for the length of the study.  By accounting for second order risk in this way, it is possible to more accurately forecast the volatility of optimized portfolios.

\section{Discussion}\label{sec:Extensions}
The techniques developed here begin to account for the costs of distributional uncertainty; however, they are not fully general, having made a number of simplifying assumptions to calculate the corrections to the risk forecast.  

Some of these assumptions may be relaxed without much difficulty.  Rather than the equal-weighted covariance matrix estimators we have considered, it is common to use an exponentially weighted estimator that puts greater weight on recent events.  For a half-life $\tau$ we may account for these effects to leading order in $K/\tau$ by replacing the number of observations $T$ with an effective time window
\bea
	T\rightarrow 2\tau/ln(2).   \label{EWMA}
\eea
Similarly, the Newey-West estimator \cite{NeweyWest} accounting for $n>1$ lags of serial correlation can be approximated with
\bea
	T\rightarrow 3T/2(n+1), \nonumber
\eea 
and for a fat-tailed process with uniform kurtosis $k$, we may take
\bea
	T\rightarrow 2T/(k-1).  \label{kurtosis}
\eea 
Note that Equation (\ref{EWMA}) accounts for the use of the EWMA estimator, but not the non-stationarity for which it is adopted.  To account for the latter, the general expression (\ref{therisk}) could be adapted to a model for the market process.  

Similarly, while the kurtosis correction (\ref{kurtosis}) approximately accounts for a uniformly kurtotic underlying process, it does not treat an optimization that attempts to minimize risk to extreme events with a utility function that penalizes assets according to their estimated tail behavior.  We expect the effects of this paper to be amplified in the context of fat-tails optimization, due to the increased estimation error associated with rare, large events.

Another generalization is to the case of constrained optimization.  We find the second order risk forecasts to be robust to the inclusion of a small number of constraints, and linear constraints may be accounted for easily.  However, it is difficult to extend analytic results to more general constraints, and suggest a heuristic based on the transfer coefficient \cite{Clark}, which measures the discrepancy between the constrained and unconstrained optimal portfolios.  A Monte Carlo approach may also be useful.

Other generalizations are more difficult.  A portfolio manager might reduce a position based on a large forecast marginal contribution to risk, creating the correlations between the portfolio weights and $\HO$ that lead to biases in the standard risk forecasts, but such an investment strategy is difficult to quantify.  More difficult still are the true \qql unknown unknowns", whose effects -- by definition -- cannot be anticipated.  

On a larger scale, second order risk provides insight into the current financial crisis, in which a financial system optimized under models and assumptions of the economy finds itself with far more risk than it had accounted for.  While our focus has been the risk of an optimized portfolio, there is likely a general principle at work.  If something has been tuned to a particular measure, that measure is likely to become exaggerated.  Similar distortions may be commonplace, with examples ranging from standardized test scores to the balance sheet information upon which executive compensation is based.

\section{Acknowledgements}

The author is grateful to Lisa Goldberg and Jose Menchero for helpful comments, and to MSCI Barra for support of this work.  A sample of related work may be found in \cite{Zhou}, \cite{Beat} and \cite{LeDoitWolf}.

\appendix
\section{Mathematical Details}

To establish (\ref{real3}), we use two identities of the Wishart distribution:
\bea
	E(\iHO)&=&\frac{\BO^{-1}}{(1-\frac{N-1}{T})}, \nonumber\\
E(\iHO\BO\iHO) &=&T^2(T-1)\left[(T-N)(T-N-1)(T-N-3)\right]^{-1}\BO^{-1} \nonumber\\
&\simeq&\left(1-\frac{N}{T}\right)^{-3}\BO^{-1} \nonumber\\
&\simeq&\left(1-\frac{N}{T}\right)^{-2}E(\HO^{-1}).\nonumber
\eea
The approximations drop $\mathcal{O}(1/N)$ and $\mathcal{O}(1/T)$ terms, for simplicity.  For the portfolio
\bea
	\Hw &=& \iHO\alpha\nonumber
\eea
comparison of the average naive forecast
\bea
E(\Bhw' \HO \Bhw)&=& \alpha' E(\iHO\HO\iHO)\alpha\nonumber\\
&=& \alpha' E(\iHO)\alpha,\nonumber
\eea
with the average true risk
\bea
E(\Bhw' \BO \Bhw)&=& \alpha' E(\iHO\BO\iHO)\alpha\nonumber\\
&=& \left(1-\frac{N}{T}\right)^{-2} \alpha' E(\iHO)\alpha,\nonumber
\eea
yields Equation (\ref{real3}).

\bibliographystyle{board}
\bibliography{all}

\end{document}